# Predicting User Roles in Social Networks using Transfer Learning with Feature Transformation

Jun Sun[1], Jérôme Kunegis[1] and Steffen Staab[1,2]
[1]Institute for Web Science and Technologies (WeST)
University of Koblenz–Landau, Germany
[2]Web and Internet Science Research Group (WAIS)
University of Southampton, UK
Email: {junsun, kunegis, staab}@uni-koblenz.de

*Abstract*—How can we recognise social roles of people, given a completely unlabelled social network? We may train a role classification algorithm on another dataset, but then that dataset may have largely different values of its features, for instance, the degrees in the other network may be distributed in a completely different way than in the first network. Thus, a way to *transfer the features of different networks to each other or to a common feature space is needed. This type of setting is called transfer learning*. In this paper, we present a transfer learning approach to network role classification based on feature transformations from each network's local feature distribution to a global feature space. We implement our approach and show experiments on real-world networks of discussions on Wikipedia as well as online forums. We also show a concrete application of our approach to an enterprise use case, where we predict the user roles in ARIS Community, the online platform for customers of Software AG, the second-largest German software vendor. Evaluation results show that our approach is suitable for transferring knowledge of user roles across networks.

## I. Introduction

Communities of people are often modelled as social networks consisting of individual actors whose roles in the community correspond to the network patterns present around their corresponding nodes. Examples of such roles for individual actors in social networks are people bridging two communities, central people through which a large part of communication passes, and outliers. In social network analysis, recognising user roles is helpful to gain deeper understanding of the underlying communities. For large online social networks, the only scalable way to achieve this is through automatic labelling of nodes, i.e. using machine learning. If, in a community, persons are already annotated with roles (by whatever method), this can be exploited to train a classifier to detect person roles in case new people appear in the community. If a role is known for only a fraction of people, then the result is a semi-supervised learning setting, for which a large number of machine learning methods has been studied. If no role labels are known however, then the role labels cannot be inferred from the dataset – an unsupervised learning algorithm could be used, but it would only output unlabelled classes without meaning attached to the classes. To overcome this, transfer learning enables one to learn knowledge from other, labelled, datasets, and apply the learnt knowledge to the unlabelled dataset [1]. In transfer learning approaches, an algorithm is trained on one dataset (the source dataset) and applied to another dataset (the target dataset). In practice, this only works as stated if the two datasets have very similar structural (and other) properties. In the general case, transfer learning approaches need to additionally specify a transfer function that maps features from one dataset to another, or to a common feature space.

In this paper, we specifically address the problem of predicting user roles in online social networks with transfer learning, based on known user role labels in other online social networks. We model our features on structural aspects of the network, which by definition are present in all social networks [2], as opposed to e.g. geolocations which are only available for some datasets. Due to the sometimes drastic differences in structural network features such as the degree distribution, we must additionally investigate specific transfer functions for such features. The result is a three-level algorithm, based on (i) the extraction of structural features of nodes, (ii) a transformation of the extracted features to a common feature space, and (iii) the classification of nodes based on the transformed structural features. In doing so, we propose a new method to transform power-law distributions, extending previous studies. The contributions of this paper are:

- We study the problem of predicting user roles in unlabelled social networks, and propose to use transfer learning to transfer knowledge from known networks. The proposed approach uses feature transformation, and experimental results show the effectiveness of it.
- We propose a method of transformation for power-law distributions. This method can be used effectively in transfer learning tasks in network analysis on features such as node degrees.

The rest of the paper is organised as follows. Section II introduces background knowledge and related work on transfer learning, network analysis and role prediction. Section III demonstrates our transfer learning–based role prediction approach in detail. Section IV shows the experimental evaluation on real datasets and a concrete use case of our approach. Section V concludes the paper.

## II. Background: Transfer Learning

In this section, we introduce the concept of transfer learning, as well as existing transfer learning algorithms for network analysis and role analysis.

## A. Transfer Learning

Traditionally, if we perform machine learning tasks such as classification, regression and clustering [3], one assumption is made: The training data from which we learn, and the test data to which we want to apply the knowledge we learnt are sampled from the same domain, i.e. they are assumed to have the same feature distribution [1]. However, this assumption might not stand true in many real scenarios, especially when we have completely new domains to work on, or completely new tasks to accomplish. In this case, transferring knowledge learnt from an existing domain is necessary. This concept of knowledge transfer is called *transfer learning*. In this paper we adopt the definition of transfer learning given by Pan and Yang [1]:

> "Given a source domain $D_S$ and learning task $T_S$, a target domain $D_T$ and learning task $T_T$, transfer learning aims to help improve the learning of the target predictive function $f_T(\,\cdot\,)$ in $D_T$ using the knowledge in $D_S$ and $T_S$, where $D_S \neq D_T$ or $T_S \neq T_T$."

We use the term *source dataset* to refer to a dataset that belongs to the source domain $D_S$, that we learn knowledge from; and *target dataset* to refer to a dataset that belongs to the target domain $D_T$, to which we want to transfer the knowledge which we have learnt from the source dataset.

In our study, both source and target datasets contain different user interaction networks, where each node is a user and each directed edge represents a user interaction. We assume the ground truth about user roles in present in the source dataset, but not in the target dataset.

## B. Transductive Transfer Learning

In [1], the term *transductive transfer learning* is defined as a special case of transfer learning, where, in contrast to inductive learning, *transductive* emphasises that the learning targets $T_T$ and $T_S$ are the same[1]:

> "Given a source domain $D_S$ and a corresponding learning task $T_S$, a target domain $D_T$ and a corresponding learning task $T_T$, transductive transfer learning aims to improve the learning of the target predictive function $f_T(\,\cdot\,)$ in $D_T$ using the knowledge in $D_S$ and $T_S$, where $D_S \neq D_T$ and $T_S = T_T$. In addition, some unlabeled target-domain data must be available at training time."

Note that in our case, the target dataset could be completely unseen during the training part (see Section III), therefore no target-domain data (labelled or unlabelled) are available at training time. This makes the task more challenging [4].

To summarise, the domains $D_S$ and $D_T$ are two different social networks. The learning tasks $T_S$ and $T_T$ are the same: both to determine user roles in the networks. The predictive function $f_T(\,\cdot\,)$ gets the structural features of a node as input, and returns the role of this node. Therefore, the task this paper focuses on can be catagorised as transductive transfer learning, if we relax the condition that "some unlabelled target-domain data must be available at training time."

---

[1]Note that the terms *transductive learning* and *transfer learning* are orthogonal, i.e. a transductive learning problem can be either a transfer learning problem or a non-transfer learning problem, and vice versa [4].

## C. Transfer Learning Algorithms for Network Analysis

Henderson et al. have proposed ReFeX[2] [5], an algorithm to extract nodes' structural features in a network. The idea of ReFeX is that an actor in a network is not only characterised by who the actor is, but also who its neighbours are, and where it is located in the network. Thus, ReFeX recursively combines nodes' local features and neighbourhood features. Evaluation has shown that ReFeX is scalable and suitable for a variety of transfer learning network analysis tasks, e.g., node classification and de-anonymisation.

In [2], Henderson et al. describe an algorithm called RolX to extract node roles in networks automatically. RolX first performs an unsupervised soft clustering for all nodes in the network using matrix decomposition (e.g., non-negative matrix factorisation [6]) on the structural features extracted by ReFeX or other algorithms. The result of the matrix decomposition, a soft clustering of nodes, is regarded as a probability-based role membership assignment of the nodes.

RolX also supports transfer learning (across-network role classification), if ground truth of user roles is present in one network (the source dataset). The probability-based role memberships of the nodes in the source dataset can serve as features to train a classifier (e.g., a logistic regression model), which can then be used to classify user roles in another network, based on the role membership features in the target dataset obtained in the same way.

## III. PROPOSED METHOD

In this section, we illustrate the detailed procedure of our transfer learning–based role prediction approach. Figure 1 shows an overview of the procedure, which can be divided into two parts: *training* (top row) and *application* (bottom row).

In both parts, we generate structural features of nodes from the corresponding network (the source network for the training part, and the target network for the application part). The feature generation processes for both parts can be done separately, because we do not assume the availability of the target dataset during the training part. In each part, with a social network $G$ as input, we generate the feature matrix $F_{n \times m}$, where $n$ is the number of nodes in $G$ and $m$ is the total number of features. Inside $F_{n \times m}$, each node (user) has $m$ structural features $x_j, (j \in \{1, 2, \ldots, m\})$, which are expected to be non-domain-specific. The feature generation process consists of three steps: (i) feature extraction, (ii) feature transfer by feature transformation, and (iii) feature aggregation. They will be explained respectively in the following subsections.

In the source dataset, we also have the ground truth of user roles in the network, which is denoted as a vector $R_S$ of length $n_S$ where $n_S$ is the size (i.e. number of nodes) of the source network. Each value $r_i \in \{1, 2, \ldots, k\}$, $(i \in \{1, 2, \ldots, n_S\})$ in $R_S$ denotes the role of the $i$-th user, where $k$ is the total number of possible roles.

---

[2]Recursive Feature eXtraction

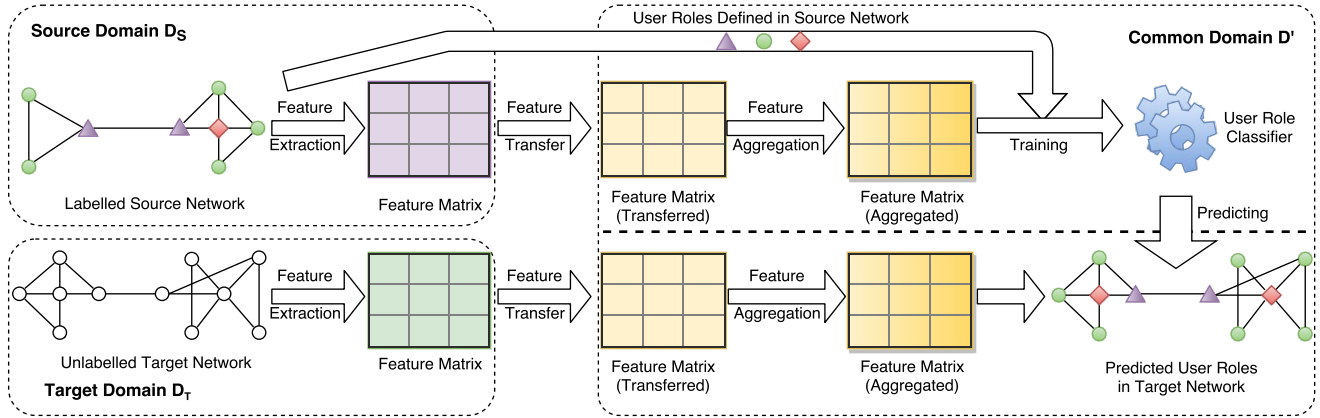

Fig. 1. Overview of the transfer learning procedure for role prediction described in this paper. We learn from the source dataset how to detect the following three (exemplary) predefined user roles: central users (diamonds), bridging users (triangles), and normal users (circles); and transfer the knowledge to the unlabelled target network to predict the user roles in it. The components within the source domain $D_S$ and within the common domain $D'$ but above the dashed line are obtained during *training*, while the ones within the target domain $D_T$ and within the common domain $D'$ but below the dashed line are obtained during *application*.

With $R_S$ and the feature matrices $F_S, F_T$ for both source and target networks, the role prediction problem reduces to a classification problem. In the training part, we optimise a predictive function $r_\text{predict} = f(x_1, x_2, \ldots, x_m)$ that maps input data (i.e. a node's features $x_1, x_2, \ldots, x_m$) into a role[3] $r_\text{predict} \in \{1, 2, \ldots, k\}$, so that $r_\text{predict}$ matches the corresponding value in $R_S$ to a certain extent [8].

Such a predictive function $f$ can be regarded as a user role classifier which classifies nodes into different roles in other networks, since the features in $F_S$ are expected to be non-domain-specific, and are already transformed in a way that they can match across networks. In the application part, we use $f$ to compute $r_\text{predict}$ for all nodes in $G_T$ in order to predict user roles in the target network.

The rest of the section describes the individual steps.

### A. Feature Extraction

In social networks, users are grouped in different social roles according to their behaviours. Users with similar behaviours belong to the same role, and different roles represent different behaviours of user groups. User behaviour is reflected in network structure, and thus we can examine the structural features of a node to analyse a user's behaviour, and further identify his or her social role [2].

Structural features of nodes can be extracted by looking at only the structure (e.g., the adjacency matrix) of the network, without requiring information on additional attributes of nodes or links (e.g., users' geolocations as node attributes, or message contents as link attributes in a user interaction network) [5]. In traditional machine learning, it is often helpful to take this additional information into consideration. However, these node and link attributes are usually domain-specific, and may not be applicable in other domains. In transfer learning, blindly transferring knowledge may not be successful, or even make the performance of learning worse [1]. On the contrary, structural features are usually common across networks. Therefore, in this study we only consider structural features.

[3]Or similarly, when performing soft or fuzzy classification [7], into a distribution of probability that the input node belongs to each role, as in our case.

Given the adjacency matrix $A$ of a network, we compute the following five structural features for each node as its *base features*:

- Degree: The total number of links of the given node.
- Indegree: The number of inlinks of the given node.
- Outdegree: The number of outlinks of the given node.
- Local clustering coefficient: The probability that a pair of neighbours of the given node are connected [9]. We ignore edge directions when we compute the local clustering coefficient for each node.
- PageRank: The stationary probability at the given node in a converged random surfing process in a network [10]. See Section III-B3 for details.

### B. Feature Transfer by Feature Transformation

The main challenge in transfer learning is that the distributions of features differ between the source and target datasets. Thus, features that are extracted from different networks are often not directly comparable. Therefore, after all base features are extracted, we transform them via different methods in order to make them comparable across networks. The transformation of features to a dataset-independent space of values is performed separately for each dataset.

Feature transformation is especially difficult since the target dataset might not be seen during the training phase [4], [11]. In our approach, the general idea of feature transformation is to define a common feature distribution for each kind of base feature, which is more likely to be comparable across networks. Therefore, the feature transformation procedure is

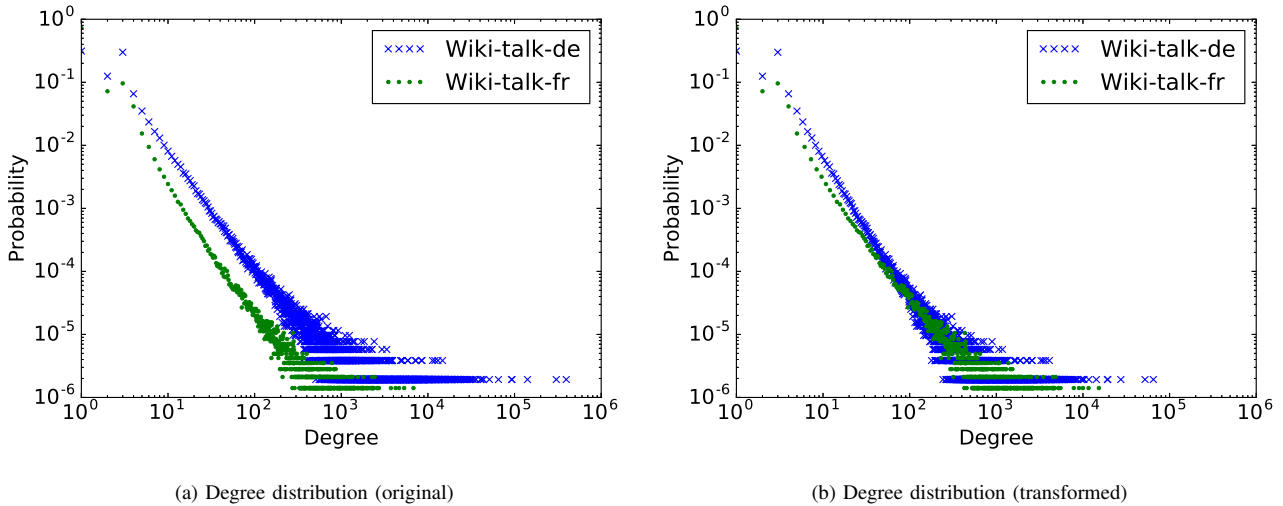

(a) Degree distribution (original)  (b) Degree distribution (transformed)

Fig. 2. Degree distributions of networks `Wiki-talk-de` and `Wiki-talk-fr` (see Section IV-A) before and after power-law degree transformation. Each dot in the plot represents the probability (Y axis) of a degree value (X axis) in the network. Two separated curves overlap after the transformation.

network-independent and order-free, i.e. we do not need to access the target network when we perform feature transformation for the source network, and vice versa.

In the following, we discuss the different transformation methods we evaluated.

*1) Quantile transformation:* There exists an intuitive approach, quantile transformation, for which we transform any given feature value into its *quantile* value [13], which is always within the range $[0, 1)$. The quantile of a feature value is defined as the probability that any value in this domain is less than this value:

$$x' = \text{quantile}(x) = P(x_j < x), j \in \{1, 2, \ldots, |X|\},$$

where $x'$ serves as the transformed feature value. For example, if the original feature values are:

$$X = [1, 0, 1, 5, 2]$$

then the transformed feature values $X'$ (quantiles) will be:

$$X' = [0.2, 0, 0.2, 0.8, 0.6]$$

Considering the definition of quantile, the transformed features always have a value within $[0, 1)$, and thus they are comparable across feature domains.

However, quantile transformation will lose information of the original feature's distribution. Here we consider a simple example: a different set of original feature values $[1, 0, 1, 5, 4]$ will also lead to the same quantiles $[0.2, 0, 0.2, 0.8, 0.6]$ as in the previous example. More generally, $X'$ will be uniformly distributed within $[0, 1)$, if there are no equal values in the original feature values $X$, regardless of the distribution in $X$.

*2) Power-law degree transformation:* For some of the base features such as nodes' degree, studies have shown that in real social networks, they follow power-law distribution approximately [14], [15]:

$$p(x) = c \cdot x^{-\alpha}$$
$$\alpha > 1, x \in [x_{\min}, +\infty), x_{\min} > 0 \quad (1)$$

Given this prior knowledge, we can plug in the idea of quantile transformation (Equation 2) and transform any kind of power-law like feature distributions into one common power-law distribution.

$$\int_{x_{\min}}^{x} p(x)dx = \int_{x'_{\min}}^{x'} p'(x')dx' \quad (2)$$

We choose the power-law distribution:

$$p'(x') = x'^{-2} \ (x' \in [1, +\infty)) \quad (3)$$

as the target distribution of transformation for the ease of calculation. Combining Equations 1, 2 and 3, we get:

$$x' = \left(\frac{x}{x_{\min}}\right)^{\alpha-1},$$

where $x'$ is the transformed feature value. Considering our scenario, degree $d$ starts from 1 in most cases, the formular can be further simplified to

$$x' = d^{\alpha-1}, \qquad (d \in [1, +\infty)).$$

In our implementation, we use the method by Clauset et al. [15] to fit a power-law distribution and estimate the exponential $\alpha$. For features that do not start from 1 such as the indegree and the outdegree, we only perform fitting and transformation on the values that fullfil power-law distribution well.

Figure 2 shows the degree distributions of two networks before and after our transformation. The original curves of degree distributions are clearly separated (as in Figure 2a), while being overlapping after the transformation (as in Figure 2b). This indicates that our transformation method can transform degree distributions from different networks into a common power-law distribution.

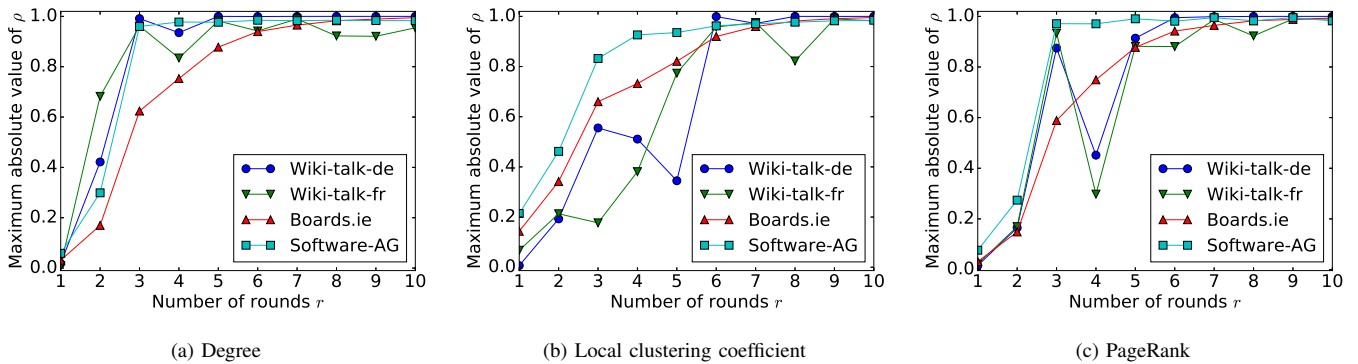

Fig. 3. Convergence procedure of aggregating neighbourhood features for (a) degree, (b) local clustering coefficient and (c) PageRank in different datasets (each curve represents a dataset). The X axis shows the round number of feature aggregation, while the Y axis shows the maximum absolute value of the pearson correlation coefficient $\rho$ between the newly generated neighbourhood feature in the current round and the corresponding old features. New features with bigger $\rho$ provide less information [12]. In most cases, $\rho$ gets larger than 0.9 after 5 rounds. We omit other datasets and features here, since they show similar patterns.

*3) PageRank transformation:* The standard PageRank of nodes in a network is defined as the stationary probability distribution in a converged random surfing process, i.e. the probability that a surfer is located at a certain node [10]. In a directed graph $G(V, E)$, the PageRank $pr(v)$ of a node $v$ is defined as:

$$pr(v) = (1-\alpha) \sum_{(u,v) \in E} \frac{pr(u)}{out(u)} + \frac{\alpha}{|V|},$$

where $\alpha$ is the random teleportation parameter, and $out(u)$ is the outdegree of $u$. PageRank is usually used to measure the centrality of nodes in directed networks. However, PageRank is not applicapable to compare nodes from different networks, because it is not independent of the network size. Berberich et al. have pointed out that, as a network gets larger, the PageRank values of nodes tend to get smaller. In order to overcome this problem, they have proposed *normalised PageRank* [16], which is defined by:

$$\hat{pr}(v) = \frac{pr(v)}{pr_{\text{low}}},$$

where $pr_{\text{low}}$ is the theoretical lower bound of the PageRank considering the random teleportation at each step of the random walk and at dangling nodes (i.e. nodes with outdegree zero):

$$pr_{\text{low}} = \frac{1}{|V|} \left( \alpha + (1-\alpha) \sum_{d \in D} pr(d) \right),$$

where $D \subseteq V$ denotes the set of dangling nodes in the network. The normalised PageRank has been proved to be independent of network size and comparable across networks [16]. Hence, we use it as a tranformation for our base feature PageRank.

### C. Feature Aggregation

It is important to notice that, inside a network, one can characterise a node not only by who it is, but also who are its neighbours, and where it is located. In terms of machine learning, we do not only consider a node's local features, but also look into its neighbourhood's features and the network structure around it. Inspired by the idea of *recursive features* proposed in [5], for each node in the network, we generate its *neighbourhood features* by aggregating its neighbours' features step by step. For more details, in the first round, for each node and each local feature, we compute the average feature value of its neighbours[4] and store it as a new feature. In the following rounds, we aggregate the features that we get in the last round in the same way.

Obviously, this repetitive progress can be done infinitely without the limitation of a round number. And if we continue this repetitive progress, the resulted feature will converge to a certain vector which is the right eigenvector of $P$ [17], thus will provide less and less information and will increase the computational overhead. Hence, we introduce a parameter $r$ to limit the number of rounds that we perform feature aggregation. In our experiments, we use $r = 5$ in order to achieve a good balance between classification performance and computational overhead (see Figure 3).

If we have the adjacency matrix $A$ of the network and the base feature matrix $F_0$, where $A_{(u,v)}$ denotes the number of directed edges from node $u$ to node $v$, and $F_{0(u,j)}$ is the $j$-th base feature value of node $u$, then there is a simple way to do the feature aggregation using matrix multiplication. Algorithm 1 shows the pseudocode for the feature aggregation progress, with $A$, $F_0$ and $r$ as input. The output $F_i$ ($i \in \{1, 2, \ldots, r\}$) is the neighbourhood feature matrix generated in round $i$. Concatenating $F_0, F_1, \ldots, F_r$ horizontally, we get the structural feature matrix $F$ for the given network.

---
[4] Two nodes are neighbours of each other if they are connected by at least one edge.

**Algorithm 1** Aggregating neighbourhood features
**Input:** $A, F_0, r$
**Output:** $F_1, F_2, \ldots, F_r$ ▷ Neighbourhood features
1: **procedure** FEATUREAGGREGATION($A, F_0, r$)
2: $\quad A \leftarrow sgn(A + A')$ ▷ Make the graph undirected and simple
3: $\quad D \leftarrow diag(A \cdot \mathbf{1}_{n \cdot 1})$
4: $\quad P \leftarrow D^{-1} \cdot A$
5: $\quad$ **for** $i \leftarrow 1, r$ **do**
6: $\quad\quad F_i \leftarrow P \cdot F_{i-1}$
7: $\quad$ **end for**
8: **end procedure**

### D. Reduction to Classification Problem

In the training part of our transfer learning procedure, once we get the feature matrix $F_S$ for the source dataset, with the labelled user roles $R_S$, the problem of user role prediction reduces to a classification problem. Hence, we are able to train a classifier that classifies nodes (users) into their correct classes (roles). In our implementation, we use a random forest classifier [18]. The parameters of the random forest classifier are set according to the method used by Oshiro and colleagues [19]. Once the classifier is trained, it is able to return the probability that each node (user) belongs to each class (roles), given the structural features of the node as input. And it can be saved to classify user roles in multiple target datasets.

In the application part, once we obtain the feature matrix $F_T$ for the target dataset, we can use the pre-trained classifier to predict the user roles in the target network.

## IV. EXPERIMENTS

In this section, we first demonstrate the datasets that we use in our experiments. We then show experiments on a set of networks to evaluate the performance of our approach, including a concrete application of our approach to a real use case.

### A. Datasets

We use the following real-world datasets in our experiments.

*a) Wiki-talk:* In Wikipedia, each registered user has a talk page that can be used for discussion. We extract the user interaction networks of all user talk pages of Wikipedia in the 28 languages with the highest number of articles (at the time of dataset creation[5]) [20]. Each language forms an individual directed network, in which each node is a Wikipedia user, and each directed edge (`User_ID_A, User_ID_B, timestamp`) represents a user interaction: User A wrote a message on User B's talk page at a certain time. Each user has an access level [21], which we interpret as the following roles:

- **Administrator.** Administrators refer to the accounts that have high level of access to contents and maintenance tools in Wikipedia. We combine the users that are granted as "sysops" or "bureaucrat" by the communities at RfA or RfB[6], and regard them as administrators in our study.
- **Bot.** Bots are used in Wikipedia for automatically or semi-automatically improving contents. Bot accounts are marked as "bot" by an administrator, and each has specific tasks that it performs [22].
- **Normal user.** Other users that are not catagorised as administrators or bots.

The proportions of both bots and administrators are very small, althought they vary highly among all sub-datasets, from 0.0027% to 5.97% and from 0% to 0.72% respectively.

*b) Boards.ie:* This dataset (denoted `Boards.ie`) contains the user interaction network of Boards.ie, one of the largest online forum of Ireland. The network consists of 66,931 nodes (users) and 6,877,447 edges. Each directed edge (`User_ID_A, User_ID_B, timestamp`) denotes that User A wrote a comment to User B's post or comment. A subset of users are annotated with the roles *Administrator*, *Moderator*, *Subscriber* and *Banned*. Table I shows the frequency and proportion of each role in the dataset.

TABLE I
META INFORMATION OF THE `Boards.ie` DATASET.

| Role name | Frequency | Proportion |
|---|---|---|
| Administrator | 6 | 0.00896% |
| Moderator | 367 | 0.548% |
| Subscriber | 124 | 0.185% |
| Banned | 9420 | 14.1% |

*c) Software AG ARIS Community:* This dataset (denoted as `Software-AG`) contains a user interaction network in Software AG's ARIS Community. The ARIS Community is the internal Business Process Management (BPM) system used in Software AG company, to enable better collaboration, engagement and sharing. At the time we extracted it, it had 9,566 threads and 20,538 comments by 4,216 users, and the total user number was 394,716. Similar to `Boards.ie`, we use a directed edge to represent a user's comment to another user's post or comment. Since its user roles are not labelled, we only use this dataset as a target dataset.

### B. Experiment on Wikipedia Networks

In this experiment, we use `Wiki-talk` to evaluate the performance of our transfer learning approach, and try to identify administrators among bots and normal users.

*1) Administrator classifier:* We use all 28 sub-datasets in `Wiki-talk` and build 28 binary classifiers for administrators respectively. Each of them is applied to all the other 27 sub-datasets. Therefore, we have 756 pairs of source and target datasets. Different feature transformation methods (see Section III-B) are used and compared:

- `None`: performing no feature transformation;

---
[5]The 28 languages are, listed alphabetically by their ISO 639 code: ar bn br ca cy de el en eo es eu fr gl ht it ja lv nds nl oc pl pt ru sk sr sv vi zh.

[6]Requests for adminship (RfA), Requests for bureaucratship (RfB)

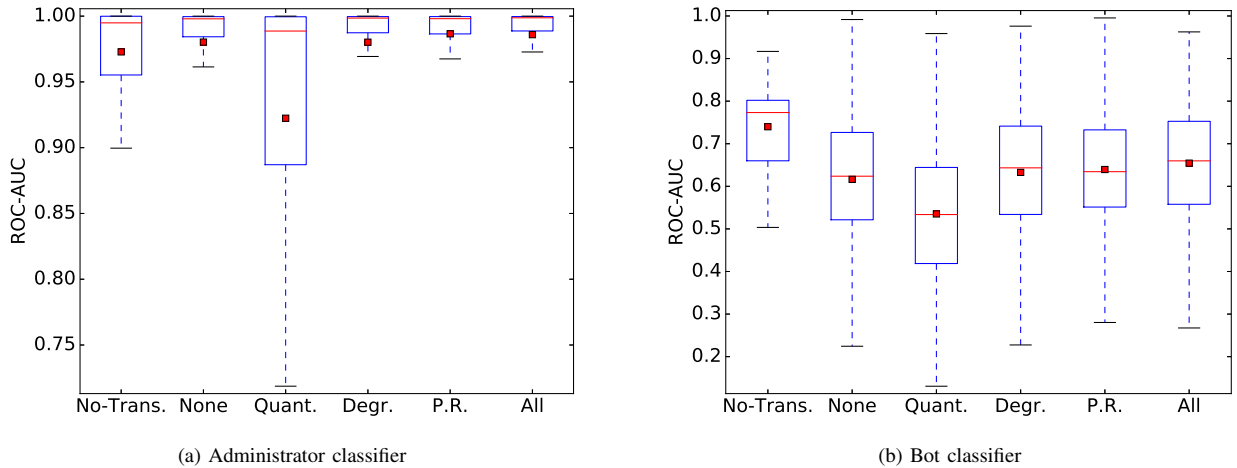

Fig. 4. ROC-AUC performance of the two classifiers in different settings (see Section IV-B). In each box plot, the red bar shows the median value, while the red dot shows the mean value of the ROC-AUC in each experiment. When we perform both power-law degree transformation and PageRank transformation (notated as `All`), the average performance is increased by 1% for the administrator classifier and 6% for the bot classifier than doing no feature transformation.

- `Quant.`: performing quantile transformation for all base features;
- `Degr.`: performing power-law degree transformation for degree, indegree and outdegree;
- `P.R.`: performing PageRank transformation for PageRank;
- `All`: combining `Degr.` and `P.R.`.

In addition, traditional machine learning (i.e. training a classifier from partial data in a network and apply the classifier to the rest data in the same network, denoted `No-Trans.`) is used as a baseline. We use the ROC-AUC metric to measure the performance of the classifiers.

As shown in Figure 4a, we can achieve high ROC-AUC when we use traditional machine learning, which means identifying administrators is an easy task. Beyond our expectation, even we do not perform any feature transformation, the performance of transfer learning is better than traditional machine learning on average. This is probably because only partial data in the network are used in training when we do traditional machine learning. Quantile transformation does not make the performance better. The reason is that it will break the internal distribution of features, as we stated in Section III-B1. Both power-law degree transformation and PageRank transformation can improve the performance, and combining both can achieve the best overall performance.

*2) Bot classifier:* Similarly, we train 28 bot classifiers in the datasets and evaluate their performance. The result is shown in Figure 4b. As we can see, the ROC-AUC is not so high even for traditional machine learning. This means identifying bots is a relatively challenging task. Besides, similar to the situation in the administrator classifier, all feature transformation methods have better performance compared to when we do not perform any feature transformation, except quantile transformation.

### C. Experiment on Forum Networks

Now we apply our transfer learning–based approach to predict user roles in `Software-AG`. We use `Boards.ie` as the source dataset, since it is also a forum user communication network. For the four annotated roles in `Boards.ie` (Administrator, Moderator, Subscriber and Banned), we build four binary classifiers separately.

Since we do not have ground truth of user roles in `Software-AG`, there is no direct way to evaluate our result. However, there is ground truth of user trustiness (around 40% users are annotated as trusted users). If we assume that there are big overlaps between trusted users and users with the three "positive" roles (Administrator, Moderator and Subscriber), then the proportions of trusted users with the three positive roles should be higher than average. Similarly, if being a trusted user and being a "Banned" user are mutually exclusive, the proportion of trusted users with the "negative" role (Banned) should be lower than average.

Hence, we use our classifiers to identify top-$k$ users in `Software-AG` with the highest probabilities of belonging to each role, and compare the identified users with the annotated trusted users as an indirect verification. The result is shown in Figure 5. As we can see, when $k$ is small (which means the probability threshold is high), the percentages of trusted users among the top-$k$ identified users with positive roles are large, while the percentage of trusted users among the top-$k$ identified "Banned" users is small. When $k$ gets larger (which means the probability threshold gets lower), the percentages converge to the baseline (40% trusted users). This indicates that our approach is able to predict users with the three positive roles (Administrator, Moderator and Subscriber), if they have big overlaps with trusted users; as well as those with the negative role (Banned), if they are exclusive with trusted users.

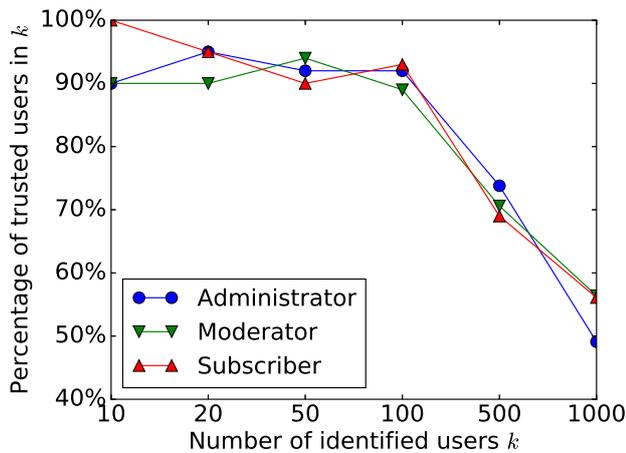
(a) Positive roles

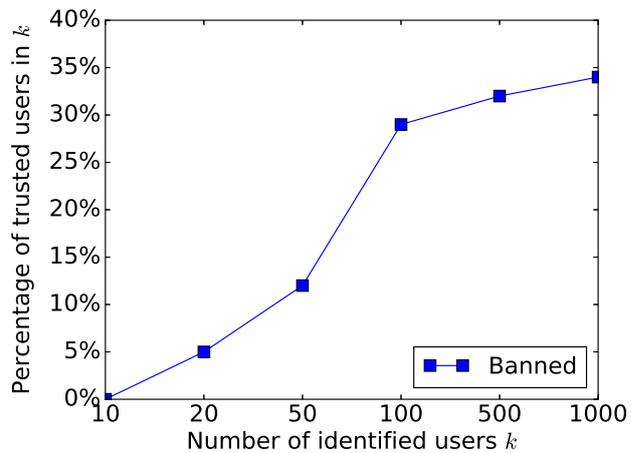
(b) Negative role

Fig. 5. Identifying top-$k$ users with largest probabilities (given by our classifier) of belonging to (a) positive roles (Administrator, Moderator and Subscriber) and (b) negative role (Banned) in `Software-AG`. The Y axis shows the percentage of trusted users among the top-$k$ identified users with the given role. With high probability thresholds (low $k$s), the percentages of trusted users in the identified top-$k$ users with positive roles (Administrator, Moderator and Subscriber) are high, while the percentage of trusted users in the identified top-$k$ "Banned" users is low. When we lower the probability thresholds to get more possible users with given roles, the percentages of trusted users converge to 40%, which is the proportion of trusted users in all users.

## V. Conclusion

In this paper, we proposed a transfer learning–based approach with feature transformation to predict user roles in social networks. Furthermore, we also proposed a method to transform power-law like distributions. Experiments on real network datasets were carried out and the results showed the effectiveness of our approach.

Future work inclues extending the method and study other problems in network analysis with transfer learning, such as link prediction and friend recommendation.

## Acknowledgment

We would like to thank Software AG for providing the dataset. The research leading to these results has received funding from the European Community's Seventh Framework Programme under grant agreement nº 610928, REVEAL.

## References

[1] S. J. Pan and Q. Yang, "A survey on transfer learning," *IEEE Transactions on knowledge and data engineering*, vol. 22, no. 10, pp. 1345–1359, 2010.
[2] K. Henderson, B. Gallagher, T. Eliassi-Rad, H. Tong, S. Basu, L. Akoglu, D. Koutra, C. Faloutsos, and L. Li, "RolX: structural role extraction & mining in large graphs," in *Int. Conf. on Knowledge Discovery and Data Mining*, 2012, pp. 1231–1239.
[3] X. Wu, V. Kumar, J. R. Quinlan, J. Ghosh, Q. Yang, H. Motoda, G. J. McLachlan, A. Ng, B. Liu, S. Y. Philip *et al.*, "Top 10 algorithms in data mining," *Knowledge and Information Systems*, vol. 14, no. 1, pp. 1–37, 2008.
[4] A. Arnold, R. Nallapati, and W. W. Cohen, "A comparative study of methods for transductive transfer learning," in *Int. Conf. on Data Mining Workshops*, 2007, pp. 77–82.
[5] K. Henderson, B. Gallagher, L. Li, L. Akoglu, T. Eliassi-Rad, H. Tong, and C. Faloutsos, "It's who you know: Graph mining using recursive structural features," in *Int. Conf. on Knowledge Discovery and Data Mining*, 2011, pp. 663–671.
[6] D. D. Lee and H. S. Seung, "Algorithms for non-negative matrix factorization," in *Advances in neural information processing systems*, 2001, pp. 556–562.
[7] E. Binaghi and A. Rampini, "Fuzzy decision making in the classification of multisource remote sensing data," *Optical Engineering*, vol. 32, no. 6, pp. 1193–1204, 1993.
[8] D. M. Hawkins, "The problem of overfitting," *J. of Inf. and Computer Sciences*, vol. 44, no. 1, pp. 1–12, 2004.
[9] D. J. Watts and S. H. Strogatz, "Collective dynamics of 'small-world' networks," *Nature*, vol. 393, no. 6684, pp. 440–442, 1998.
[10] L. Page, S. Brin, R. Motwani, and T. Winograd, "The PageRank citation ranking: Bringing order to the Web," 1999.
[11] D. Koschützki, K. A. Lehmann, D. Tenfelde-Podehl, and O. Zlotowski, "Advanced centrality concepts," in *Network Analysis*, 2005, pp. 83–111.
[12] M. A. Hall, "Correlation-based feature selection for machine learning," Ph.D. dissertation, The University of Waikato, 1999.
[13] R. J. Serfling, *Approximation theorems of mathematical statistics*. John Wiley & Sons, 2009, vol. 162.
[14] M. E. Newman, "Clustering and preferential attachment in growing networks," *Physical Review E*, vol. 64, no. 2, p. 025102, 2001.
[15] A. Clauset, C. R. Shalizi, and M. E. Newman, "Power-law distributions in empirical data," *SIAM review*, vol. 51, no. 4, pp. 661–703, 2009.
[16] K. Berberich, S. Bedathur, G. Weikum, and M. Vazirgiannis, "Comparing apples and oranges: normalized pagerank for evolving graphs," in *Int. Conf. on World Wide Web*, 2007, pp. 1145–1146.
[17] R. Mises and H. Pollaczek-Geiringer, "Praktische Verfahren der Gleichungsauflösung," *Zeitschrift für Angewandte Mathematik und Mechanik*, vol. 9, no. 2, pp. 152–164, 1929.
[18] A. Liaw and M. Wiener, "Classification and regression by randomforest," *R news*, vol. 2, no. 3, pp. 18–22, 2002.
[19] T. M. Oshiro, P. S. Perez, and J. A. Baranauskas, "How many trees in a random forest?" in *Int. Workshop on Machine Learning and Data Mining in Pattern Recognition*, 2012, pp. 154–168.
[20] J. Sun and J. Kunegis, "Wiki-talk datasets," Apr. 2016. [Online]. Available: http://dx.doi.org/10.5281/zenodo.49561
[21] Wikipedia, "User access levels — Wikipedia, the free encyclopedia," 2016, [Accessed 13-July-2016]. [Online]. Available: https://en.wikipedia.org/w/index.php?title=Wikipedia:User_access_levels&oldid=727508916
[22] ——, "Bot policy — Wikipedia, the free encyclopedia," 2016, [Accessed 25-July-2016]. [Online]. Available: https://en.wikipedia.org/w/index.php?title=Wikipedia:Bot_policy&oldid=730456542